\documentclass[aps,prl,groupedaddress,twocolumn,showpacs,longbibliography]{revtex4-2}
\usepackage{amsmath}
\usepackage{amsthm, amssymb}    
\usepackage{braket}              
\usepackage{graphicx}
\usepackage{color}
\usepackage{times, verbatim}      
\usepackage{mathtools}
\usepackage{natbib}
\usepackage{hyperref}
\hypersetup{
       colorlinks=true,
}
\usepackage{etoolbox}
\newtoggle{includeHubbardDetails}
\togglefalse{includeHubbardDetails}

\begin{document}

\title{
Many Body Scars as a Group Invariant Sector of Hilbert Space
}

\author{{K. Pakrouski$^{1}$, P.N. Pallegar$^{1}$,  F.K. Popov$^{1}$, I.R. Klebanov$^{1,2}$}         }
\affiliation{$^{1}$Department of Physics, Princeton University, Princeton, NJ 08544}
\affiliation{$^{2}$Princeton Center for Theoretical Science, Princeton University, Princeton, NJ 08544}

\begin{abstract}
We present a class of Hamiltonians $H$ for which a sector of the Hilbert space invariant under a Lie group $G$, which is not a symmetry of $H$, possesses the essential properties of many-body scar states. These include the absence of thermalization and the ``revivals" of special initial states in time evolution.
A particular class of examples concerns interacting spin-1/2 fermions on a lattice consisting of $N$ sites (it includes deformations of the Fermi-Hubbard model as special cases), and we show that it contains two families of $N+1$ scar states. One of these families, which was found in recent literature, is comprised of the well-known $\eta$-pairing states. 
We find another family of scar states which is $U(N)$ invariant.
Both families and most of the group-invariant scar states produced by our construction in general, give rise to the off-diagonal long range order which survives at high temperatures and is insensitive to the details of the dynamics. Such states could be used for reliable quantum information processing because the information is stored non-locally, and thus cannot be easily erased by local perturbations. In contrast, other scar states we find are product states which could be easily prepared experimentally. 
The dimension of scar subspace is directly controlled by the choice of group $G$ and can be made exponentially large.
 \end{abstract}

\date{\today}


\maketitle

The concept of many-body scar states has recently emerged as a novel type of weak ergodicity breaking~\cite{Shiraishi2017ScarsConstruction,Turner_2018,Moudgalya:2018,AbaninScarsSU2Dynamics,Khemani:2019vor,Sala_2020,Prem:2018,Schecter:2019oej,SciPostPhys.3.6.043,IadecolaHubbardAlmostNUPRL2019,Shibata:2020yek,michailidis2020stabilizing,2020arXiv200413800M,PRLPapicClockModels,VedikaScarsVsIntegr,Pal2020ScarsFromFrustration,mark2020unified,iadecola2020quantum,moudgalya2020etapairing,Magnifico:2019kyj}. These states are typically found in the bulk of the spectrum and thus play a role at high temperatures. The scars are special because they have low (area-law) entanglement entropy, do not thermalize, and lead to the exact ``revivals" of the initial state of the system initialized with scars. Therefore, the information stored in the system does not dissipate at finite temperature, holding promise for potential applications in quantum information processing.

The current knowledge of the nature of this phenomenon is based on the identification of scars in a variety of systems, such as interacting fermionic models \cite{Shiraishi2017ScarsConstruction,Prem:2018,SciPostPhys.3.6.043,IadecolaHubbardAlmostNUPRL2019},
the AKLT spin chain~\cite{Moudgalya:2018},  the spin-1 XY model \cite{Schecter:2019oej}, frustrated spin systems \cite{Pal2020ScarsFromFrustration}, and a spin-$\frac{1}{2}$ domain-wall conserving model ~\cite{mark2020unified,iadecola2020quantum}. In some cases \cite{SciPostPhys.3.6.043,moudgalya2020etapairing,2020arXiv200413800M}, the scar states
are related to the well-known $\eta$-pairing states of the Hubbard model, which form a family under the $SU(2)$ symmetry called pseudospin \cite{etaPairingYang89,yang1990so,ZhangHubbardSO41991}.
There has been experimental observation of the approximate revivals~\cite{RydbergExperimentRevivals}, yet a general understanding of the underlying structures leading to the existence of scars is not yet available.

The Hamiltonians exhibiting scars can be often brought to the form $H=H_0 + H_1$, such that $H_1$ breaks some of the symmetries of $H_0$ and has
a special property that it annihilates a subsector of the Hilbert space $\mathbb{S}$ consisting of eigenstates of $H_0$. In this paper, we discuss how the symmetry properties of the Hilbert space can be used to construct scars systematically. We analyze a rich class of models where the scar subsector $\mathbb{S}$ is invariant under the action of a continuous group $G$, which is bigger than the symmetry of the full Hamiltonian. We will show
(see SM Sec. II) that the requisite hermitian operator $H_1$ must have the form $H_1 = \sum_a O_a T_a$, where $T_a$ are generators of the symmetry group $G$ and $O_a$ is any operator s.t. the product $O_a T_a$ is Hermitian. For $H_0$, the simplest option is that it has symmetry $G$, i.e. $[H_0, T_a]=0$, but the most general condition is that 
\begin{gather}
\label{eq:H0CommCasimir}
[H_0, C^2_G]= W\cdot C^2_G,
\end{gather}  
 where $W$ is some operator and $C^2_G$ is the quadratic Casimir of the group $G$. Then the states invariant under $G$ are eigenstates of $H_0$.

Our main finding is that for any Hamiltonian of the form
\begin{gather}
\label{eq:H0PlusOTForm}
H = H_0+\sum_a O_a T_a
\end{gather}  
the dynamics of the scar subsector $\mathbb{S}$ is governed by $H_0$ and is decoupled from the rest of the spectrum controlled by $H$. If the ergodic properties of $H_0$ and $H$ are sufficiently different every state in the decoupled sector $\mathbb{S}$ will not thermalize with the rest of the system and will thus violate the eigenstate thermalization hypothesis (ETH) \cite{deutsch1991quantum,srednicki1994chaos,rigol2008thermalization}. Because of the decoupling, the unitary time evolution starting from a state in the invariant sector cannot mix it with the rest of the system. In addition, if the energy gaps between the states from the invariant subsector have a common divisor  \footnote{this happens, for example, when the energies of all states in $\mathbb{S}$ are integers in some units}, then the unitary time evolution of a state from the invariant sector will exhibit revivals: the initial state will return to itself after equal time intervals. Therefore the states in $\mathbb{S}$ possess all of the defining properties of the many-body scar states. To our knowledge, such general constructions have not been discussed previously, and we present 
their concrete examples.

The general class of models we study includes the famous Fermi-Hubbard model and its deformations. In this context we show that, in addition to the family of states which transform as
spin $N/2$ under the pseudospin symmetry (the $\eta$-pairing states), which were recently shown to be scar states in \cite{SciPostPhys.3.6.043,2020arXiv200413800M,moudgalya2020etapairing}, there is another family of scar states. This
second family, whose states may be explicitly written down as (\ref{stateNu}) or SM (12), is invariant under the $U(N)$ symmetry which acts on the degrees of freedom on all $N$ lattice sites; it forms a multiplet of spin
$N/2$ under the $SU(2)$ which is the physical rotational symmetry in the Fermi-Hubbard and related models.

\emph{Hilbert space structure}---It is central to our approach and determines the existence and properties of the invariant subspace $\mathbb{S}$. We will focus on the Hilbert space of $M$ fermionic oscillators 
$\{c_{I},c^\dagger_{I'} \} = \delta_{I  I'} , \quad I, I'=1, \ldots,  M\ , $ 

which has dimension $2^M$ and is obviously acted on by $U(M)$. The Hilbert space forms a spinor representation of $O(2M)$ which acts on the $2M$ Majorana fermions, and we can choose $G$ to be any of its subgroups. The choice of $G$ provides an important handle on the dimension of the scar subspace: the smaller the group $G$, the bigger the invariant scar sector $\mathbb{S}$. In particular, scar sectors which are exponentially large in $M$ can be achieved for sufficiently small groups.

\begin{figure}
	\centering
	\includegraphics[scale=0.5]{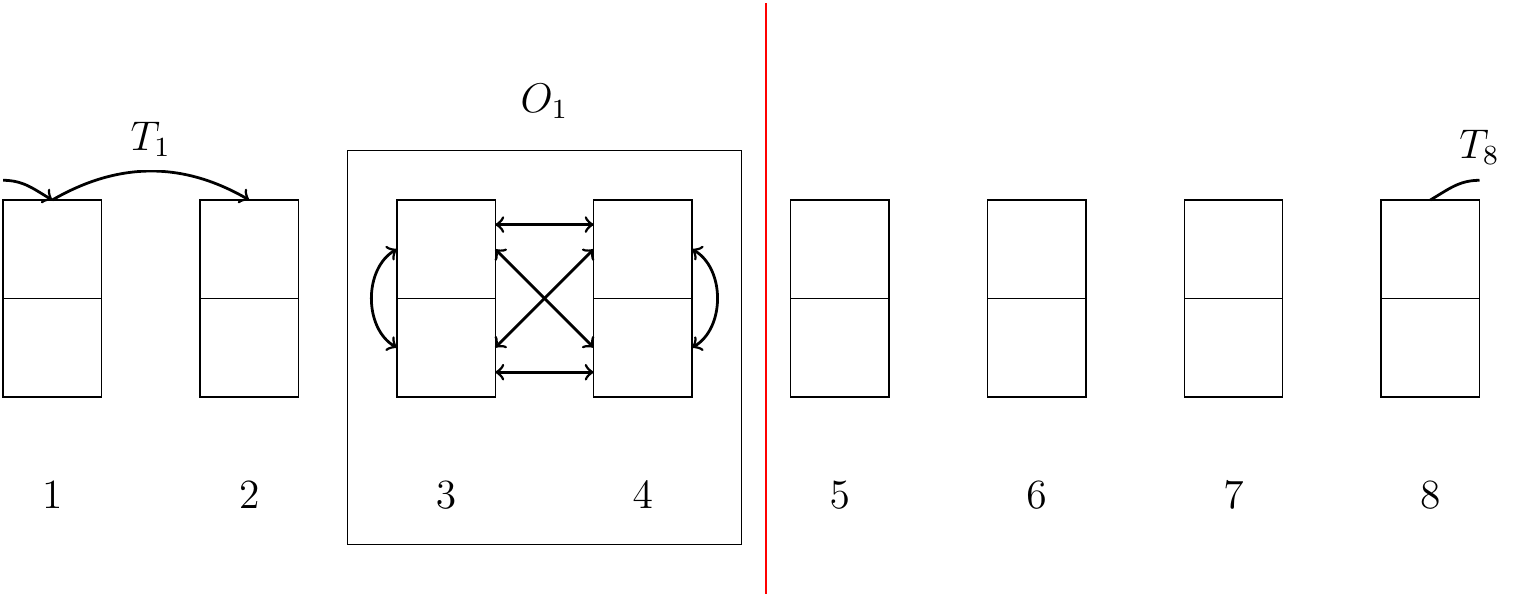}
	\caption{Schematic representation of the Hilbert space and model \eqref{eq:vectorFullH}. Each line corresponds to a hopping $T_i$ or some bilinear operator in terms of fermion operators. The red line shows the position of cut for the entanglement calculations.}
		\label{fig:LatticeView}
\end{figure}

For $M=2N$ one may interpret \cite{PhysRevB.100.075101} this Hilbert space as that of a lattice model with $N$ sites and two fermionic degrees of freedom per site (they may correspond to the two states of a spin-$1/2$ fermion).
The Hilbert space and the structure of the invariant subspace $\mathbb{S}$ we consider are thus identical to that in a number of spin-1/2 models, such as Hubbard, Hirsch and their deformations. There are two useful ways this Hilbert space can be factorized \cite{Klebanov:2018nfp,Gaitan:2020zbm}: according to the representations of $U(N)\times SU(2)$ or according to the representations of $O(N)\times SO(4) = O(N) \times SU(2)\times SU(2)$ (the relevance of group $SO(4)$ was noted long ago in the context of Hubbard model \cite{yang1990so,ZhangHubbardSO41991}). An analog of a Cauchy identity \cite{Gaitan:2020zbm} leads to the representations of the product groups being related to each other (see SM Fig. 2 for a schematic view).
Thus, choosing a singlet representation of $U(N)$ fixes the $(N+1)$-dimensional spin-$N/2$ representation of $SU(2)$ - the maximal representation of $SU(2)$ in the given Hilbert space.
An analogous relation for orthogonal groups indicates that the $O(N)$ singlets transform in the $(N/2, 0)+ (0,N/2)$ representation of $SO(4)\sim SU(2)\times SU(2)$, where we labeled the $SU(2)$ representations by their spin $J$.
Thus, there are {\it two} sets of $N+1$  $O(N)$ invariant states; each one is invariant under one of the $SU(2)$ groups and transforms as spin $N/2$ under the other.

We see that there are two natural choices for the subgroup of $U(2N)$: $G=U(N)$ or $G=O(N)$, both acting on the degrees of freedom on all $N$ lattice sites. The lattice may be thought of as one-dimensional, as in fig.~\ref{fig:LatticeView}, but the specific way the $U(N)$ or $O(N)$ indices are mapped to spatial lattice indices is not important for the purposes of finding scars. In particular, the lattice can be of arbitrary dimension, frustrated, and can have any boundary conditions.

The hopping term on this lattice is $T= \sum \limits_{aa',\sigma} t_{aa'}\, c^\dagger_{a\sigma}c_{a'\sigma}$, where the first index of $c_{a\sigma}$ labels the sites, the second index the ``spin" and $t_{aa'}$ is the hopping strength hermitian matrix. One can see that, for a general complex $t_{aa'}$, the hopping $T$ is a generator of $SU(N)$ that acts on the indices $a$ (see \cite{Klebanov:2018nfp} and SM Sec. II). 
Adding the charge $Q$ to the set of generators we would obtain generators of $U(N)$. 
For purely imaginary $t_{aa'}$ the hopping $T$ is a generator of $SO(N)$, and for real $t_{aa'}$ the situation depends on the parity of $N$ (see SM Sec. II).

\emph{Example Hamiltonian.}---Following the prescription \eqref{eq:H0PlusOTForm} we first have to choose $H_0$ that will control the scar subsector; it must satisfy \eqref{eq:H0CommCasimir}. We will use the following integrable fermionic Hamiltonian \cite{Klebanov:2018nfp}

\begin{gather} 
\label{eq:H0}
H_0 =2 \left (c^\dagger_{a\sigma} c^\dagger_{a\sigma'} c_{a'\sigma} c_{a'\sigma'} - c^\dagger_{a\sigma} c^\dagger_{a'\sigma} c_{a \sigma'} c_{a'\sigma'}\right )\notag \\ 
+2 (2- N)Q+ N (N-2)\ , \label{eq:H0} \\
\{c_{a\sigma},c^\dagger_{a'\sigma'} \} = \delta_{aa'}\delta_{\sigma\sigma'} , \quad a=1, \ldots, N\ , \  \sigma=1, \ldots, 2\ , \notag
\end{gather}
where summation over repeated indices is implied.\footnote{
A more general version of $H_0$ \cite{Klebanov:2018fzb} is considered in Sec. I of the SM where the choice of a smaller group $G$ leads to the dimension of the scar subsector growing exponentially 
in $M=N_1N_2$.} It may be viewed as a generalized Hubbard interaction term which has a continuous symmetry $O(N)\times O(2)$, in addition to the usual $U(1)$ symmetry with conserved charge $Q=\frac 1 2 [c^\dagger_{a\sigma}, c_{a\sigma} ]$. It is a special case, $N_2=N_3=2$, of the $O(N_1)\times O(N_2)\times O(N_3)$ fermionic tensor model \cite{Klebanov:2016xxf,Klebanov:2018nfp}. While in general the tensor model is not integrable \cite{Pakrouski:2018jcc}, for $N_3=2$ it is \cite{Klebanov:2018nfp,Gaitan:2020zbm}, and all of the energies are integer.  
Because of the $O(N)\times O(2)$ symmetry we have $[H_0, C^2_{U(N)}]= W\cdot C^2_{U(N)}\ne0$ and $[H_0, C^2_{O(N)}]=0$ in agreement with \eqref{eq:H0CommCasimir} which ensures the group-invariant states from $\mathbb{S}$ are eigenstates of this $H_0$.

The singlets in $\mathbb{S}$ have several quantum numbers \cite{Gaitan:2020zbm}, which can be used to distinguish them (none of them are conserved by the full Hamiltonian \eqref{eq:vectorFullH}). This includes particle number $Q$ and one of the $SU(2)$ charges
\begin{equation} 
Q_2= - i \left ( c^\dagger_{a1} c_{a2} - c^\dagger_{a2} c_{a1}\right ) = c^\dagger_{a\sigma}\sigma^2_{\sigma\sigma'} c_{a\sigma'}\ .
\end{equation}
To control the energies of the singlets and the period of revivals, we can add $\alpha Q+ \beta Q_2$ to $H_0$.

Adding also the hopping terms results in the Hamiltonian $H_T=T+ H_0 + \alpha Q+ \beta Q_2$ which remains integrable. If $T$ is a generator of $SO(N)$ (imaginary amplitude) the problem can be solved analytically and $T$ simply splits each of the $O(N)$ representations analogously to Zeeman splitting. For a complex hopping amplitude ($T$ is generator of $SU(N)$) the model is integrable in terms of level statistics but can not be simply solved analytically.

The full example Hamiltonian we will study numerically ($N=8$, $\alpha=\beta=1$, periodic boundary conditions) reads\footnote{The Hamiltonian $H=H_0+OT$ has the same properties with respect to the presence of the many-body scar states.}

\begin{gather} \label{eq:vectorFullH}
H= H_T+ 4\sum \limits^N_{a=1} O^T_a T_a + 32 \sum \limits^N_{a=1} O^Q_a (Q_a-1), \text{   where}\\
O^T_{a} =\sum_{\mathclap{a_{1,2}=(a+2),\sigma,\sigma'}}^{(a+3)} \left[ q^1_{a_{1,2},\sigma,\sigma'} c^\dagger_{ a_1 \sigma} c^\dagger_{a_2 \sigma'}  + q^2_{a_{1,2},\sigma,\sigma'}c^\dagger_{a_1 \sigma} c_{a_2 \sigma'}  + \text{h.c.}\right],  \notag \\
O^Q_{a} = \sum_{\mathclap{\sigma=1}}^{2} \left[ c^\dagger_{ a+1 \sigma} c^\dagger_{ a-1 \sigma} + \text{h.c.} \right]\ , \quad Q_a=\sum_{\sigma=1}^{2} c^\dagger_{a \sigma} c_{a \sigma}\ . \notag
\end{gather}
$T_a=8e^{i\sqrt{2}\pi} \sum_{\sigma} c^\dagger_{a\sigma}c_{a+1\sigma}$ is a translation-invariant nearest-neighbour hopping and a generator of $SU(N)$. The operator $O^T_a$ acts on sites $a+2$ and $a+3$ which ensures that $O^T_a T_a$ is hermitian and local. 
The coefficients $q^{1,2}_{a_{1,2},\sigma,\sigma'}$ are random complex numbers and this choice of the (arbitrary according to \eqref{eq:H0PlusOTForm}) operator $O$ is intended to break the symmetries of $H_0$ and to make the bulk of the spectrum ergodic. Operators $(Q_a-1)$ complement $T_a$ to form the set of $U(N)$ generators and the full Hamiltonian is of the form \eqref{eq:H0PlusOTForm} for $G=U(N)$.

\begin{figure}
	\centering
	\includegraphics[width=0.49\columnwidth]{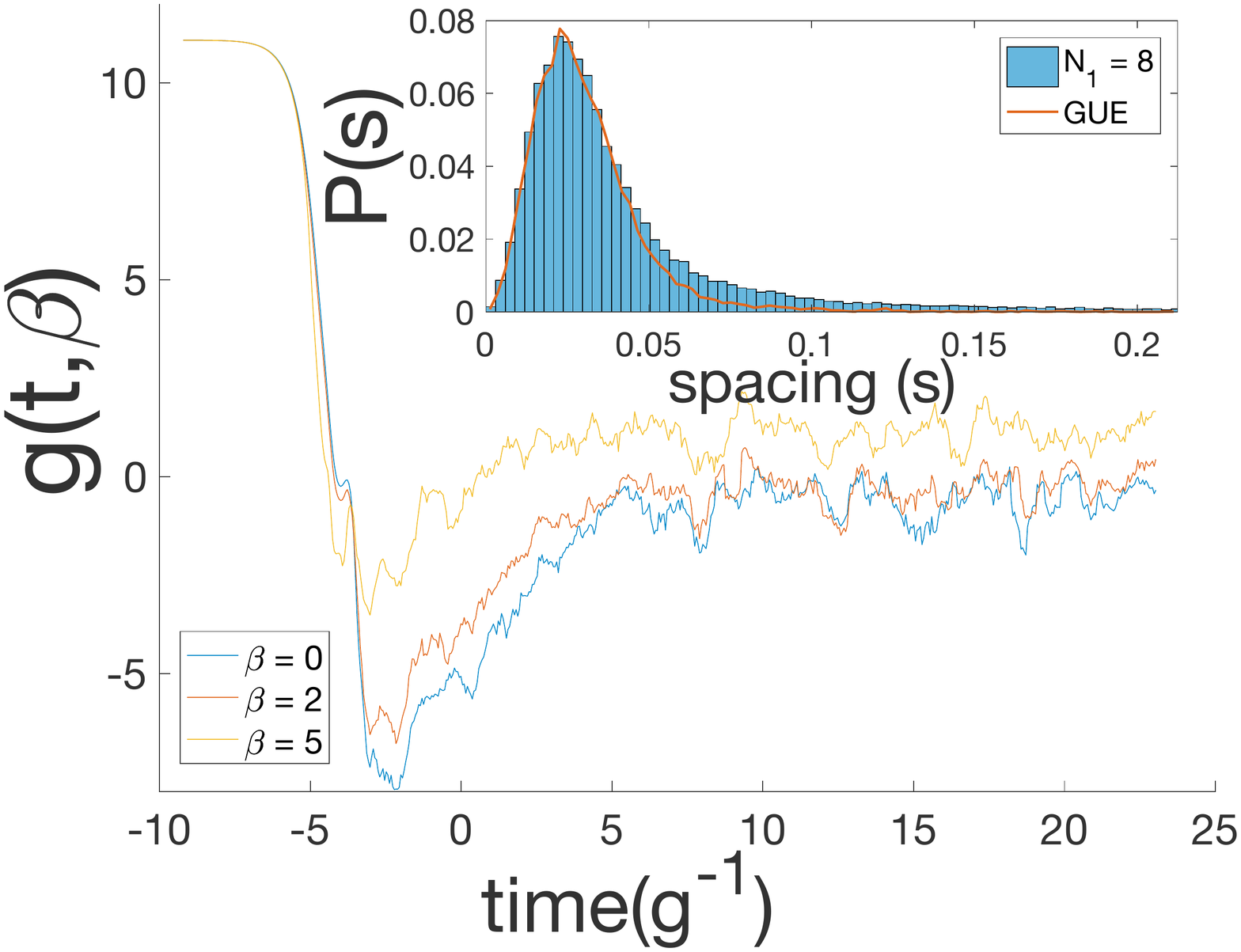}
	\includegraphics[width=0.49\columnwidth]{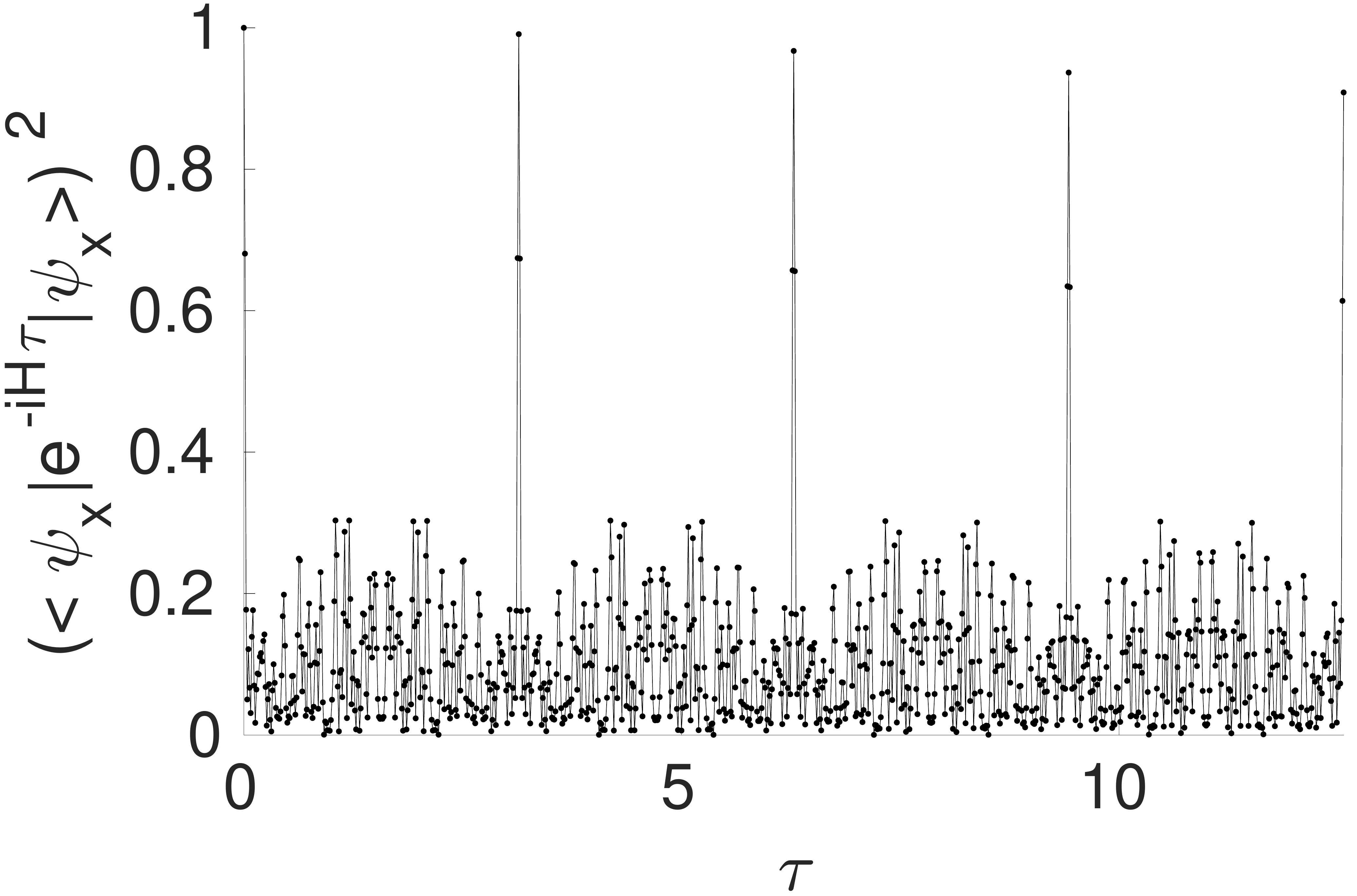}
	
	\caption{Left Panel: Histogram of the nearest neighbor eigenvalue spacings (inset, shown for the even $Q$ sector) and the spectral form factor (shown for the full spectrum) for the model in \eqref{eq:vectorFullH}. Right Panel: Time dependence of the fidelity $f(\tau)$ for vector model with $N=8$. The initial state is a linear combination of 50 eigenstates of $H$ dominated by 9 singlet states with total weight $w=\sum\limits ^{9}_{n=1} |c_n|^2=0.95$. The fidelity demonstrates oscillations with the period $T\approx3.14$ and amplitude $A\approx w^2$. The initial state composition is detailed in SM Sec V.} \label{fig:822chaos}
\end{figure}

\emph{Numerical results}---Most states in the Hilbert space will be mixed by the randomness built into $O$ while the effective Hamiltonian for the $U(N)$-invariant states in $\mathbb{S}$ remains $H_{\mathbb{S}}=H_0$. The only remaining symmetry of $H$ relates the sectors with odd and even $Q$, both described by the gaussian unitary ensemble (GUE) (see fig.~\ref{fig:822chaos} for the exact numerical results). The time-reversal symmetry is broken by the operator $c^\dagger_{a\sigma} c_{a\sigma}$  in $O_a$.

\newcommand{\Tr}{\operatorname{Tr}}

The probability distribution $P(r_k)$ of the level spacings (inset of fig.~\ref{fig:822chaos}) agrees well with the GUE overlay. It contains information about the correlation functions of close eigenvalues, whereas the spectral form factor, $g(t,\beta) = |\Tr(e^{-\beta H - i H t} )|^2 /\Tr(e^{-\beta H})^2$, also contains information about longer range correlations. The main elements of the spectral form factor (SFF) for a random matrix is a dip ramp plateau structure (for a discussion of their physics, see \cite{Cotler:2016fpe}). The presence of this structure in our system is another evidence of quantum chaos and ergodicity in its bulk spectrum. 

\begin{figure}[t!]
	\centering
	\includegraphics[width=0.23\textwidth]{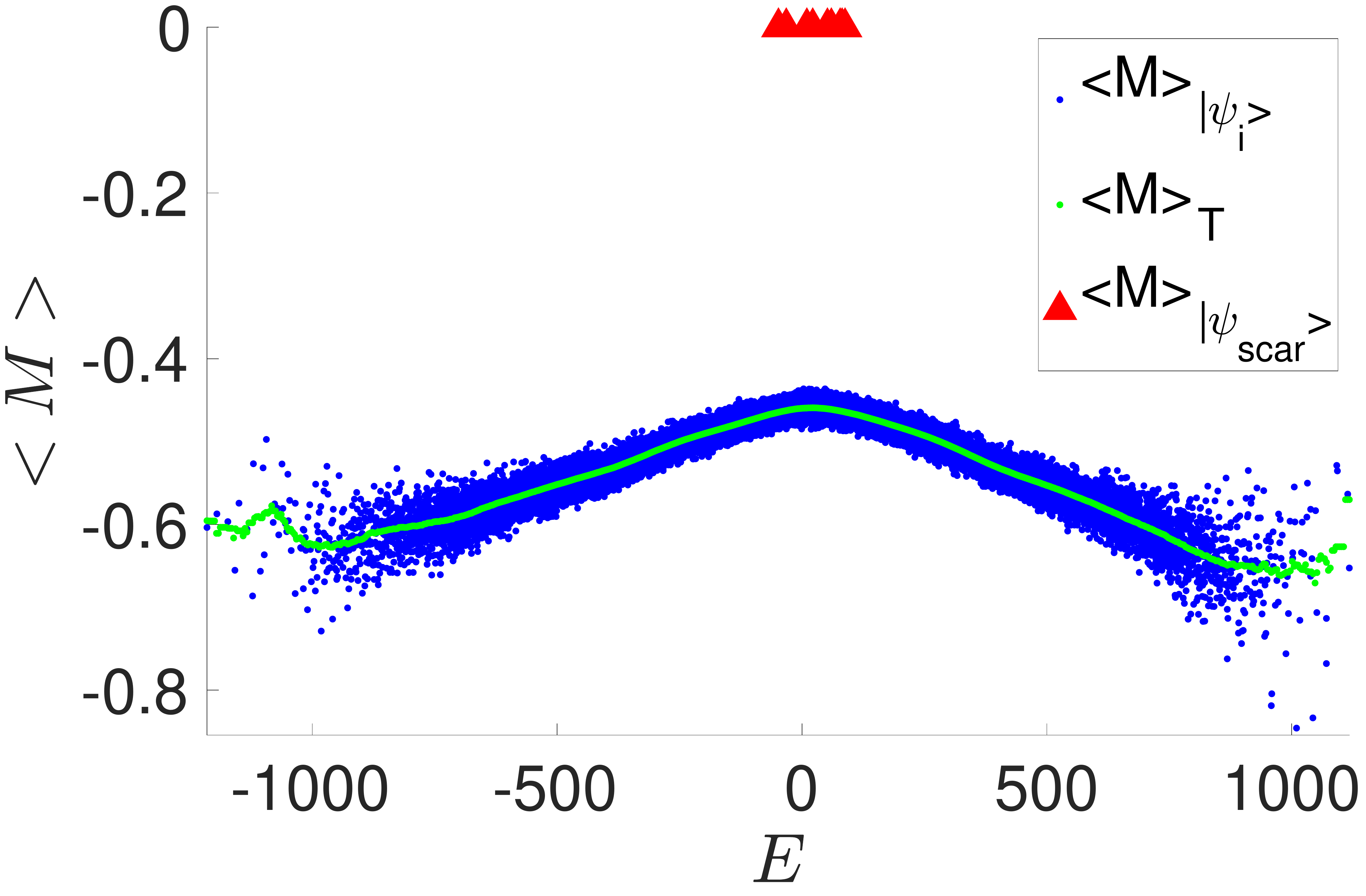}
	\hspace{2mm}
	\includegraphics[width=0.23\textwidth]{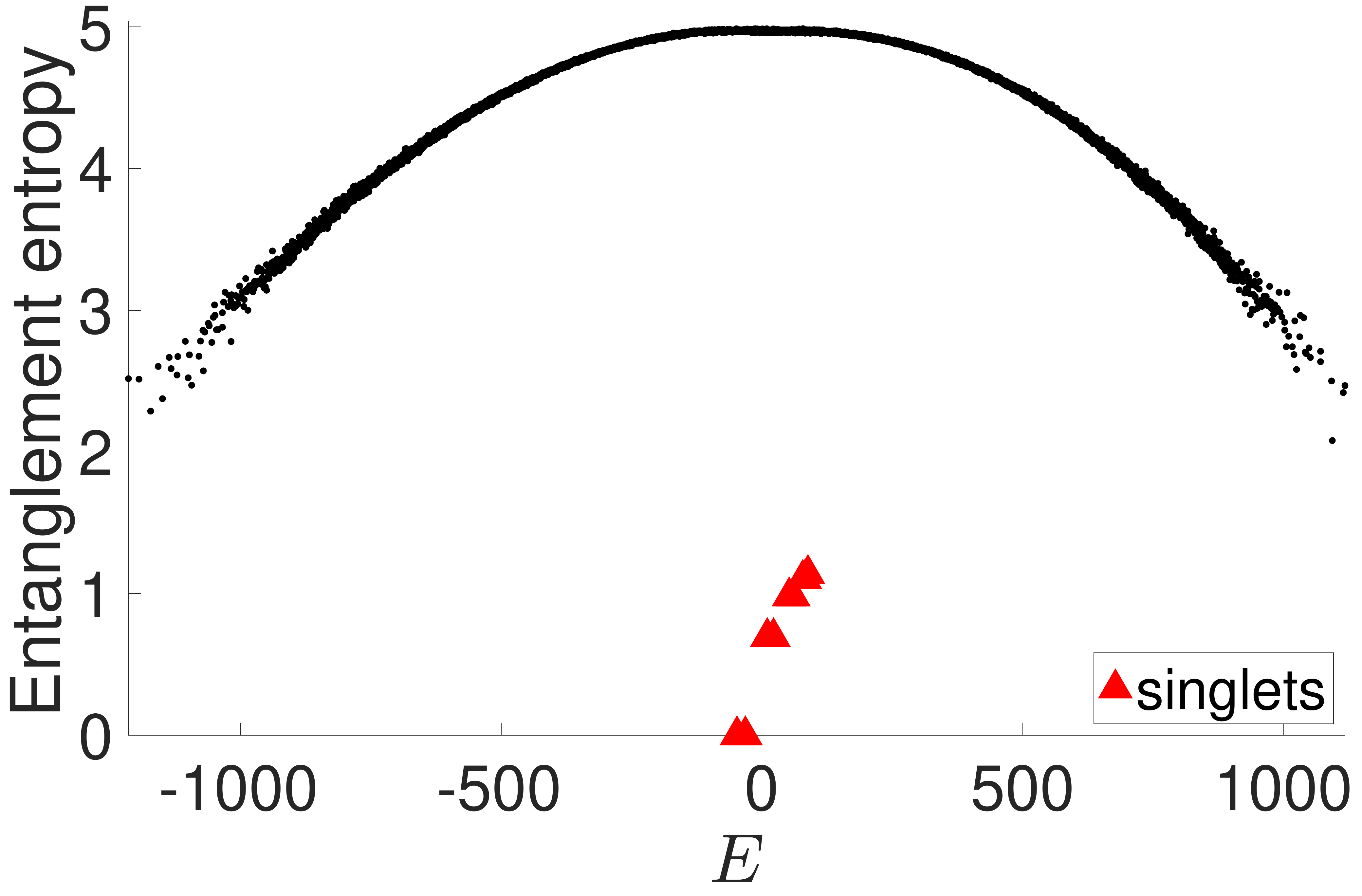}
 	\caption{Left panel: Eigenstate (blue dots) and window-averaged (green line) expectation values for $\mathcal{M} = -2 c^\dagger_{11}c_{11} c^\dagger_{12}  c_{12}$. $SU(8)$-singlet states are shown in red triangles. Right panel: Entanglement entropy calculated for every eigenstate of \eqref{eq:vectorFullH}. The cut is made between spatial sites in the middle of the chain marked by the red line in fig.~\ref{fig:LatticeView}.
	}
	\label{fig:822ETH}
\end{figure}

A more detailed characterization of ergodicity is provided in the left panel of fig.~\ref{fig:822ETH} where we test the eigenstate thermalization hypothesis (ETH), which conjectures that for any measurable local operator $\mathcal{M}$, its expectation value in an eigenstate must be approximately the same as the window-average over the nearby states at the same energy. We observe that the conjecture holds for most states in the spectrum while it is clearly violated for the nine $U(8)$ singlet states $ \left\{ \ket{n_U} \right\} $ that do not thermalize. The situation when the bulk of the spectrum is ergodic while a subset of states is not, corresponds by definition to the violation of the strong formulation of ETH (the weak formulation allows for a few ``outlier" states)~\cite{deutsch1991quantum,srednicki1994chaos,rigol2008thermalization}.

The singlet states violating strong ETH also clearly stand out in the time evolution. Consider two initial states, $\psi^s_0$, made exclusively of singlet eigenstates of $H$ and $\psi^g_0$, composed of the same number of generic states: $\ket{\psi^{s/g}_{0}} = \sum c^{s/g}_n \ket{\psi_{n}}$, where $\ket{\psi_{n}}$ is an eigenstate with energy $E_n$. The squared projection of the time-evolved state on the initial wavefunction is $f(\tau) = |\braket{\psi_0|e^{-iHt}|\psi_0}|^2 = \sum_{n,m} |c_n c_m|^2 e^{- i (E_n - E_m) \tau}$. It should quickly go to zero for generic states without correlations between $E_n$ (see SM fig. 6). A vanishing overlap with the initial state indicates that the information stored initially has fully dissipated through thermalization. This phenomenon is closely related to the dip seen in the SFF in the left panel of fig.~\ref{fig:822chaos}.

For the singlet states, all of the energies $E_n$ are integer, which means that there exists a (greatest) common divisor for all of the energy gaps between singlet states:  $\omega=\gcd(E_n-E_m)$. After the time $t^k_r=k\frac{2\pi}{\omega}, \, k\in \mathbb{Z}$ all of the exponents in $f(\tau)$ are equal to 1. This constructive interference results in ``revivals" of the (information stored in the) initial state with period $t^1_r$. 

In our example $\omega= 2$, and thus we observe the revivals with period $\pi$, as shown in the right panel of fig.~\ref{fig:822chaos}. Note that, in this calculation 5 percent of generic states were admixed to the initial state, but ideal revivals to $f(\tau)=1$ would be observed otherwise. The higher-frequency ``revivals" with smaller amplitude are due to the energy differences that are shared only by a subset of the singlet states. The energies of singlets are controlled by $H_0 + \alpha Q + \beta Q_2$ and the revivals period is a function of the parameters $\alpha$ and $\beta$. 
While a pure state is coherently oscillating in our case, an interesting construction of environment-assisted, non-stationary dynamics for mixed states was discussed in the literature recently~\cite{BucaNature2019}.

\emph{Scar subspace}---The entanglement entropy of the group-invariant states is noticeably lower compared to the generic states at the same energy as shown in the right panel of fig.~\ref{fig:822ETH}. 

Two of them are tensor products of Bell-like states formed on each site: 
\begin{gather}
\label{eq:S12}
\ket{S_1} =\bigotimes_a \frac{\ket{0_{a1} 1_{a2}} + i \ket{1_{a1} 0_{a2}}}{\sqrt{2}} = \prod_a \frac{c^\dagger_{a1} + i c^\dagger_{a2}}{\sqrt{2}} \ket{0}\\
\ket{S_2} =\bigotimes_{a} \frac{\ket{0_{a1} 1_{a2}} - i \ket{1_{a1} 0_{a2}}}{\sqrt{2}}= \prod_a \frac{c^\dagger_{a1} - i c^\dagger_{a2}}{\sqrt{2}} \ket{0}\ , \notag
\end{gather}
These states can be easily created in experiment and we provide the corresponding gate sequences in the Sec.~III of SM. The energies of these states are given by $E_{S_1}=E_0 + \alpha N + \beta N = N(\alpha + \beta + 2)$ and $E_{S_2}=E_0 + \alpha N - \beta N = N(2+\alpha-\beta)$. Initializing a system to an arbitrary superposition of these two states may be the most accessible experimental demonstration of revivals.

The complete set of the $N+1$ $U(N)$-invariant states can be constructed by acting repeatedly on the state $\ket{S_1}$ with the 
bilinear operator $\zeta= c^\dagger_{a \sigma}(\sigma^3- i \sigma^1)_{\sigma \sigma'} c_{a \sigma'}$ (this is a ``rotated" version of the zeta-operator in \cite{ZhangHubbardSO41991}):
\begin{gather}
\ket{n_U} = \frac{\zeta^n}{2^n \sqrt{\frac{N! n!}{(N-n)!}}} \ket{S_1}\ , \label{stateNu}
\end{gather}
with $n=0, \ldots, N$. Their energies are $E^{n_U}_n = N(N+2) - (N-2n)^2 + \alpha N + \beta(N-2n)$.

Another basis for this family of states is given in SM (12).
One can see that these states have the maximal possible spin $N/2$ with respect to the index $\sigma$, i.e. they transform as a $(N+1)$-dimensional representation of 
$SU(2)_\sigma$, which is the physical spin in the Fermi-Hubbard model. There is only one family which has the maximal spin. Consequently, it is quite robust under the action of any perturbation that preserves this spin.
 Namely, {\it any spin-preserving perturbation} will map this representation to itself which means these states will continue to violate strong ETH while the revivals may disappear as a result of changing their energies. 

If instead of a complex hopping strength we chose a purely imaginary one (or a real hopping strength and bipartite lattice), then the hopping terms are generators of $SO(N)$.
As it was explained above, the Hilbert space may be decomposed \cite{Gaitan:2020zbm} according to representations of $O(N)\times SO(4)$. 
There are $2N+2$ $O(N)$ singlet states that could be organized in two sets. One of these sets is $\ket{n_U}$, for which the $O(N)\times SU(2)$ symmetry is further enhanced to $U(N)$. The other set of $N+1$ states is 
\begin{gather}
\ket{n_O} = \frac{\eta^n}{\sqrt{\frac{N! n!}{(N-n)!}}} \ket{0} \  , \qquad \eta = \sum_{a=1}^N c^\dagger_{a1}c^\dagger_{a2}\ ,
\label{epair}
\end{gather}
with $n=0, \ldots, N$.
These states are invariant under $G=O(N)\times SU(2)$. They are equivalent to the exact eigenstates of the Hubbard model originally identified using the celebrated $\eta$-pairing~\cite{etaPairingYang89} and recently demonstrated to be many-body scar states~\cite{SciPostPhys.3.6.043,moudgalya2020etapairing} (to obtain (\ref{epair}) we need to transform from the real hopping amplitude used in 
\cite{etaPairingYang89} to our imaginary one, see SM Sec II).

Let us emphasize that the Hamiltonian $H$ does not respect all the symmetries possessed by
the two scar sectors. In particular, it breaks translation invariance while both $\ket{n_O}$ and $\ket{n_U}$ are manifestly invariant under lattice translations. Thus, the scars appear in the enhanced symmetry sectors of Hilbert space, in accordance with our general arguments.

It can also be shown (to appear as a separate publication) that the Fermi-Hubbard and Heisenberg Hamiltonians in arbitrary dimension can be written in the form \eqref{eq:H0PlusOTForm}; thus, the appearance of many-body scar states in these models is a special case of our construction. As a consequence, the states $\ket{n_U}$ are eigenstates and scar states in the (extended) Hubbard model, and other spin-1/2 models. In table~\ref{tab:invSubSpStruct} we 
summarize the properties of the scar subspace in the $D$-dimensional spin-1/2 models of the form~\eqref{eq:H0PlusOTForm}.

The off-diagonal long range order (ODLRO) has been linked in literature to the high-Tc superconductivity~\cite{etaPairingYang89,yang1962concept} and was shown to be present in the $\ket{n_O}$ scars~\cite{etaPairingYang89}. In the $\ket{n_U}$ scars, the ODLRO is most naturally characterized by the correlator $G_{U} = \braket{s|c^\dagger_{i1} c_{i2} c^\dagger_{j2} c _{j1} |s}$, which does not depend on the coordinates $i$ and $j$ of the sites for any $\ket{s} \in \mathbb{S}_U$ (see SM Sec IV).

\begin{table}[t!]
	\caption{ Structure of the invariant subspace $\mathbb{S}$ in spin-1/2 lattice models depending on the hopping amplitude $t$. See SM Sec II for derivation and more detailed discussion.}	
	\label{tab:invSubSpStruct}
	\begin{center}
		\begin{tabular}{|c|c|c|c|c|}
			\hline
							& real $t$										& imaginary $t$ 								& complex $t$ \\
			\hline
			odd $N$			& $\braket{\left\{\ket{n_U}\right\}}$							& $\braket{\left\{\ket{n_U}\right\} \cup \left\{\ket{n_O}\right\}}$	 	& $\braket{\left\{\ket{n_U}\right\}}$ \\
			\hline
			even $N$			& $\braket{\left\{\ket{n_U}\right\} \cup \left\{\ket{n_O}'\right\}}$ 	&  $\braket{\left\{\ket{n_U}\right\} \cup \left\{\ket{n_O}\right\}}$		& $\braket{\left\{\ket{n_U}\right\}}$ \\
			\hline
		\end{tabular}
	\end{center}
\end{table}

\emph{Discussion}---The presence of group invariant scar states $\mathbb{S}$ is a property of a Hilbert space once the conditions on the Hamiltonian outlined in this paper are satisfied. This universality explains why the scar states identified to date in different models with the same Hilbert space can be identical.

The group-invariance requirement is non-local, and the resulting scar states are invariant under lattice translations. As a consequence, the degrees of freedom on all of the sites become entangled which spreads the information over the whole system. This leads to the relative insensitivity of group-invariant states to local perturbations and protection of the quantum information ~\cite{Milekhin:2020zpg}. 
The invariant scar states do not thermalize and form a closed, decoherence-free subspace, where non-commuting transformations can act and universal quantum computation may be performed~\cite{DecohFreeQComputation1998Chuang}.

This combination of properties makes the group-invariant scar states an interesting platform for robust quantum information processing.

The gauge/gravity duality \cite{Maldacena:1997re,Gubser:1998bc,Witten:1998qj} 
is a set of correspondences between conventional gauged models without gravity and higher-dimensional gravitational systems. 
In quantum mechanical models, gauge fields are non-dynamical, so the gauging is equivalent to truncation of the Hilbert space to a group-invariant sector 
\cite{Gross:1990ub,Gross:1990md,Witten:2016iux,Klebanov:2016xxf,Klebanov:2018nfp,Maldacena:2018vsr}.

It would be interesting to explore possible connections between the group-invariant scars and gauge/gravity duality.

The broad framework we presented allows for constructing a model that could be simply realized in experiment. An example Hamiltonian could be $B_y Q_2  + \sum_a \left(M_a+M_{a+1}\right)  
 \left( c^\dagger_{a+1\sigma}c_{a\sigma} + \text{h.c.} \right)$, where $M_a = c^\dagger_{a1}c_{a1} - c^\dagger_{a2}c_{a2}$ is the magnetization at site $a$. 
 The eigenstates include the translationally invariant $\ket{n_U}$ scars including $\ket{S_1}$ with $E_{S_1}=B_yN$ and $\ket{S_2}$ with $E_{S_2}=-B_yN$.

\emph{Acknowledgements}---We are grateful to A. Prem, D. Abanin, A. Bernevig, D. Calugaru, A. Dymarsky, M. Gullans, A. Milekhin, S. Moudgalya, and G. Tarnopolsky for valuable discussions.
We also thank A. Bernevig, S. Moudgalya and B. Bu{\v c}a for comments on a draft of this paper. 
The simulations presented in this work were performed on computational resources managed and supported by Princeton's Institute for Computational Science $\&$ Engineering and OIT Research Computing.
This research was supported in part by the US NSF under Grants No.~PHY-1620059 and PHY-1914860.  
K.P. was also supported by the Swiss National Science Foundation through the Early Postdoc.Mobility Grant No. P2EZP2$\_$172168 and by DOE grant No. de-sc0002140.

\bibliography{conciseScarsPaper}

\begin{thebibliography}{48}%
\makeatletter
\providecommand \@ifxundefined [1]{%
 \@ifx{#1\undefined}
}%
\providecommand \@ifnum [1]{%
 \ifnum #1\expandafter \@firstoftwo
 \else \expandafter \@secondoftwo
 \fi
}%
\providecommand \@ifx [1]{%
 \ifx #1\expandafter \@firstoftwo
 \else \expandafter \@secondoftwo
 \fi
}%
\providecommand \natexlab [1]{#1}%
\providecommand \enquote  [1]{``#1''}%
\providecommand \bibnamefont  [1]{#1}%
\providecommand \bibfnamefont [1]{#1}%
\providecommand \citenamefont [1]{#1}%
\providecommand \href@noop [0]{\@secondoftwo}%
\providecommand \href [0]{\begingroup \@sanitize@url \@href}%
\providecommand \@href[1]{\@@startlink{#1}\@@href}%
\providecommand \@@href[1]{\endgroup#1\@@endlink}%
\providecommand \@sanitize@url [0]{\catcode `\\12\catcode `\$12\catcode
  `\&12\catcode `\#12\catcode `\^12\catcode `\_12\catcode `\%12\relax}%
\providecommand \@@startlink[1]{}%
\providecommand \@@endlink[0]{}%
\providecommand \url  [0]{\begingroup\@sanitize@url \@url }%
\providecommand \@url [1]{\endgroup\@href {#1}{\urlprefix }}%
\providecommand \urlprefix  [0]{URL }%
\providecommand \Eprint [0]{\href }%
\providecommand \doibase [0]{https://doi.org/}%
\providecommand \selectlanguage [0]{\@gobble}%
\providecommand \bibinfo  [0]{\@secondoftwo}%
\providecommand \bibfield  [0]{\@secondoftwo}%
\providecommand \translation [1]{[#1]}%
\providecommand \BibitemOpen [0]{}%
\providecommand \bibitemStop [0]{}%
\providecommand \bibitemNoStop [0]{.\EOS\space}%
\providecommand \EOS [0]{\spacefactor3000\relax}%
\providecommand \BibitemShut  [1]{\csname bibitem#1\endcsname}%
\let\auto@bib@innerbib\@empty
\bibitem [{\citenamefont {Shiraishi}\ and\ \citenamefont
  {Mori}(2017)}]{Shiraishi2017ScarsConstruction}%
  \BibitemOpen
  \bibfield  {author} {\bibinfo {author} {\bibfnamefont {N.}~\bibnamefont
  {Shiraishi}}\ and\ \bibinfo {author} {\bibfnamefont {T.}~\bibnamefont
  {Mori}},\ }\bibfield  {title} {\bibinfo {title} {Systematic construction of
  counterexamples to the eigenstate thermalization hypothesis},\ }\href
  {https://doi.org/10.1103/PhysRevLett.119.030601} {\bibfield  {journal}
  {\bibinfo  {journal} {Phys. Rev. Lett.}\ }\textbf {\bibinfo {volume} {119}},\
  \bibinfo {pages} {030601} (\bibinfo {year} {2017})}\BibitemShut {NoStop}%
\bibitem [{\citenamefont {Turner}\ \emph {et~al.}(2018)\citenamefont {Turner},
  \citenamefont {Michailidis}, \citenamefont {Abanin}, \citenamefont {Serbyn},\
  and\ \citenamefont {Papić}}]{Turner_2018}%
  \BibitemOpen
  \bibfield  {author} {\bibinfo {author} {\bibfnamefont {C.~J.}\ \bibnamefont
  {Turner}}, \bibinfo {author} {\bibfnamefont {A.~A.}\ \bibnamefont
  {Michailidis}}, \bibinfo {author} {\bibfnamefont {D.~A.}\ \bibnamefont
  {Abanin}}, \bibinfo {author} {\bibfnamefont {M.}~\bibnamefont {Serbyn}},\
  and\ \bibinfo {author} {\bibfnamefont {Z.}~\bibnamefont {Papić}},\
  }\bibfield  {title} {\bibinfo {title} {Weak ergodicity breaking from quantum
  many-body scars},\ }\href {https://doi.org/10.1038/s41567-018-0137-5}
  {\bibfield  {journal} {\bibinfo  {journal} {Nature Physics}\ }\textbf
  {\bibinfo {volume} {14}},\ \bibinfo {pages} {745–749} (\bibinfo {year}
  {2018})}\BibitemShut {NoStop}%
\bibitem [{\citenamefont {Moudgalya}\ \emph {et~al.}(2018)\citenamefont
  {Moudgalya}, \citenamefont {Regnault},\ and\ \citenamefont
  {Bernevig}}]{Moudgalya:2018}%
  \BibitemOpen
  \bibfield  {author} {\bibinfo {author} {\bibfnamefont {S.}~\bibnamefont
  {Moudgalya}}, \bibinfo {author} {\bibfnamefont {N.}~\bibnamefont
  {Regnault}},\ and\ \bibinfo {author} {\bibfnamefont {B.~A.}\ \bibnamefont
  {Bernevig}},\ }\bibfield  {title} {\bibinfo {title} {Entanglement of exact
  excited states of affleck-kennedy-lieb-tasaki models: Exact results,
  many-body scars, and violation of the strong eigenstate thermalization
  hypothesis},\ }\href {https://doi.org/10.1103/PhysRevB.98.235156} {\bibfield
  {journal} {\bibinfo  {journal} {Phys. Rev. B}\ }\textbf {\bibinfo {volume}
  {98}},\ \bibinfo {pages} {235156} (\bibinfo {year} {2018})}\BibitemShut
  {NoStop}%
\bibitem [{\citenamefont {Choi}\ \emph {et~al.}(2019)\citenamefont {Choi},
  \citenamefont {Turner}, \citenamefont {Pichler}, \citenamefont {Ho},
  \citenamefont {Michailidis}, \citenamefont {{Papi\ifmmode \acute{c}\else
  {\'c}\fi{}}}, \citenamefont {Serbyn}, \citenamefont {Lukin},\ and\
  \citenamefont {Abanin}}]{AbaninScarsSU2Dynamics}%
  \BibitemOpen
  \bibfield  {author} {\bibinfo {author} {\bibfnamefont {S.}~\bibnamefont
  {Choi}}, \bibinfo {author} {\bibfnamefont {C.~J.}\ \bibnamefont {Turner}},
  \bibinfo {author} {\bibfnamefont {H.}~\bibnamefont {Pichler}}, \bibinfo
  {author} {\bibfnamefont {W.~W.}\ \bibnamefont {Ho}}, \bibinfo {author}
  {\bibfnamefont {A.~A.}\ \bibnamefont {Michailidis}}, \bibinfo {author}
  {\bibfnamefont {Z.}~\bibnamefont {{Papi\ifmmode \acute{c}\else {\'c}\fi{}}}},
  \bibinfo {author} {\bibfnamefont {M.}~\bibnamefont {Serbyn}}, \bibinfo
  {author} {\bibfnamefont {M.~D.}\ \bibnamefont {Lukin}},\ and\ \bibinfo
  {author} {\bibfnamefont {D.~A.}\ \bibnamefont {Abanin}},\ }\bibfield  {title}
  {\bibinfo {title} {Emergent su(2) dynamics and perfect quantum many-body
  scars},\ }\href {https://doi.org/10.1103/PhysRevLett.122.220603} {\bibfield
  {journal} {\bibinfo  {journal} {Phys. Rev. Lett.}\ }\textbf {\bibinfo
  {volume} {122}},\ \bibinfo {pages} {220603} (\bibinfo {year}
  {2019})}\BibitemShut {NoStop}%
\bibitem [{\citenamefont {Khemani}\ and\ \citenamefont
  {Nandkishore}(2020)}]{Khemani:2019vor}%
  \BibitemOpen
  \bibfield  {author} {\bibinfo {author} {\bibfnamefont {V.}~\bibnamefont
  {Khemani}}\ and\ \bibinfo {author} {\bibfnamefont {R.}~\bibnamefont
  {Nandkishore}},\ }\bibfield  {title} {\bibinfo {title} {{Local constraints
  can globally shatter Hilbert space: a new route to quantum information
  protection}},\ }\href {https://doi.org/10.1103/PhysRevB.101.174204}
  {\bibfield  {journal} {\bibinfo  {journal} {Phys. Rev. B}\ }\textbf {\bibinfo
  {volume} {101}},\ \bibinfo {pages} {174204} (\bibinfo {year} {2020})},\
  \Eprint {https://arxiv.org/abs/1904.04815} {arXiv:1904.04815
  [cond-mat.stat-mech]} \BibitemShut {NoStop}%
\bibitem [{\citenamefont {Sala}\ \emph {et~al.}(2020)\citenamefont {Sala},
  \citenamefont {Rakovszky}, \citenamefont {Verresen}, \citenamefont {Knap},\
  and\ \citenamefont {Pollmann}}]{Sala_2020}%
  \BibitemOpen
  \bibfield  {author} {\bibinfo {author} {\bibfnamefont {P.}~\bibnamefont
  {Sala}}, \bibinfo {author} {\bibfnamefont {T.}~\bibnamefont {Rakovszky}},
  \bibinfo {author} {\bibfnamefont {R.}~\bibnamefont {Verresen}}, \bibinfo
  {author} {\bibfnamefont {M.}~\bibnamefont {Knap}},\ and\ \bibinfo {author}
  {\bibfnamefont {F.}~\bibnamefont {Pollmann}},\ }\bibfield  {title} {\bibinfo
  {title} {Ergodicity breaking arising from hilbert space fragmentation in
  dipole-conserving hamiltonians},\ }\bibfield  {journal} {\bibinfo  {journal}
  {Physical Review X}\ }\textbf {\bibinfo {volume} {10}},\ \href
  {https://doi.org/10.1103/physrevx.10.011047} {10.1103/physrevx.10.011047}
  (\bibinfo {year} {2020})\BibitemShut {NoStop}%
\bibitem [{\citenamefont {Moudgalya}\ \emph {et~al.}(2019)\citenamefont
  {Moudgalya}, \citenamefont {Prem}, \citenamefont {Nandkishore}, \citenamefont
  {Regnault},\ and\ \citenamefont {Bernevig}}]{Prem:2018}%
  \BibitemOpen
  \bibfield  {author} {\bibinfo {author} {\bibfnamefont {S.}~\bibnamefont
  {Moudgalya}}, \bibinfo {author} {\bibfnamefont {A.}~\bibnamefont {Prem}},
  \bibinfo {author} {\bibfnamefont {R.}~\bibnamefont {Nandkishore}}, \bibinfo
  {author} {\bibfnamefont {N.}~\bibnamefont {Regnault}},\ and\ \bibinfo
  {author} {\bibfnamefont {B.~A.}\ \bibnamefont {Bernevig}},\ }\bibfield
  {title} {\bibinfo {title} {Thermalization and its absence within krylov
  subspaces of a constrained hamiltonian},\ }\href@noop {} {\  (\bibinfo {year}
  {2019})},\ \Eprint {https://arxiv.org/abs/1910.14048} {arXiv:1910.14048
  [cond-mat.str-el]} \BibitemShut {NoStop}%
\bibitem [{\citenamefont {Schecter}\ and\ \citenamefont
  {Iadecola}(2019)}]{Schecter:2019oej}%
  \BibitemOpen
  \bibfield  {author} {\bibinfo {author} {\bibfnamefont {M.}~\bibnamefont
  {Schecter}}\ and\ \bibinfo {author} {\bibfnamefont {T.}~\bibnamefont
  {Iadecola}},\ }\bibfield  {title} {\bibinfo {title} {{Weak Ergodicity
  Breaking and Quantum Many-Body Scars in Spin-1 XY Magnets}},\ }\href
  {https://doi.org/10.1103/PhysRevLett.123.147201} {\bibfield  {journal}
  {\bibinfo  {journal} {Phys. Rev. Lett.}\ }\textbf {\bibinfo {volume} {123}}
  (\bibinfo {year} {2019})},\ \Eprint {https://arxiv.org/abs/1906.10131}
  {arXiv:1906.10131 [cond-mat.str-el]} \BibitemShut {NoStop}%
\bibitem [{\citenamefont {Vafek}\ \emph {et~al.}(2017)\citenamefont {Vafek},
  \citenamefont {Regnault},\ and\ \citenamefont
  {Bernevig}}]{SciPostPhys.3.6.043}%
  \BibitemOpen
  \bibfield  {author} {\bibinfo {author} {\bibfnamefont {O.}~\bibnamefont
  {Vafek}}, \bibinfo {author} {\bibfnamefont {N.}~\bibnamefont {Regnault}},\
  and\ \bibinfo {author} {\bibfnamefont {B.~A.}\ \bibnamefont {Bernevig}},\
  }\bibfield  {title} {\bibinfo {title} {{Entanglement of Exact Excited
  Eigenstates of the Hubbard Model in Arbitrary Dimension}},\ }\href
  {https://doi.org/10.21468/SciPostPhys.3.6.043} {\bibfield  {journal}
  {\bibinfo  {journal} {SciPost Phys.}\ }\textbf {\bibinfo {volume} {3}},\
  \bibinfo {pages} {043} (\bibinfo {year} {2017})}\BibitemShut {NoStop}%
\bibitem [{\citenamefont {Iadecola}\ and\ \citenamefont {\ifmmode
  \check{Z}\else \v{Z}\fi{}nidari\ifmmode~\check{c}\else
  \v{c}\fi{}}(2019)}]{IadecolaHubbardAlmostNUPRL2019}%
  \BibitemOpen
  \bibfield  {author} {\bibinfo {author} {\bibfnamefont {T.}~\bibnamefont
  {Iadecola}}\ and\ \bibinfo {author} {\bibfnamefont {M.}~\bibnamefont
  {\ifmmode \check{Z}\else \v{Z}\fi{}nidari\ifmmode~\check{c}\else
  \v{c}\fi{}}},\ }\bibfield  {title} {\bibinfo {title} {Exact localized and
  ballistic eigenstates in disordered chaotic spin ladders and the
  fermi-hubbard model},\ }\href
  {https://doi.org/10.1103/PhysRevLett.123.036403} {\bibfield  {journal}
  {\bibinfo  {journal} {Phys. Rev. Lett.}\ }\textbf {\bibinfo {volume} {123}},\
  \bibinfo {pages} {036403} (\bibinfo {year} {2019})}\BibitemShut {NoStop}%
\bibitem [{\citenamefont {Shibata}\ \emph {et~al.}(2020)\citenamefont
  {Shibata}, \citenamefont {Yoshioka},\ and\ \citenamefont
  {Katsura}}]{Shibata:2020yek}%
  \BibitemOpen
  \bibfield  {author} {\bibinfo {author} {\bibfnamefont {N.}~\bibnamefont
  {Shibata}}, \bibinfo {author} {\bibfnamefont {N.}~\bibnamefont {Yoshioka}},\
  and\ \bibinfo {author} {\bibfnamefont {H.}~\bibnamefont {Katsura}},\
  }\bibfield  {title} {\bibinfo {title} {{Onsager's Scars in Disordered Spin
  Chains}},\ }\href {https://doi.org/10.1103/PhysRevLett.124.180604} {\bibfield
   {journal} {\bibinfo  {journal} {Phys. Rev. Lett.}\ }\textbf {\bibinfo
  {volume} {124}},\ \bibinfo {pages} {180604} (\bibinfo {year} {2020})},\
  \Eprint {https://arxiv.org/abs/1912.13399} {arXiv:1912.13399 [quant-ph]}
  \BibitemShut {NoStop}%
\bibitem [{\citenamefont {Michailidis}\ \emph {et~al.}(2020)\citenamefont
  {Michailidis}, \citenamefont {Turner}, \citenamefont {Papi{\' c}},
  \citenamefont {Abanin},\ and\ \citenamefont
  {Serbyn}}]{michailidis2020stabilizing}%
  \BibitemOpen
  \bibfield  {author} {\bibinfo {author} {\bibfnamefont {A.~A.}\ \bibnamefont
  {Michailidis}}, \bibinfo {author} {\bibfnamefont {C.~J.}\ \bibnamefont
  {Turner}}, \bibinfo {author} {\bibfnamefont {Z.}~\bibnamefont {Papi{\' c}}},
  \bibinfo {author} {\bibfnamefont {D.~A.}\ \bibnamefont {Abanin}},\ and\
  \bibinfo {author} {\bibfnamefont {M.}~\bibnamefont {Serbyn}},\ }\bibfield
  {title} {\bibinfo {title} {Stabilizing two-dimensional quantum scars by
  deformation and synchronization},\ }\href@noop {} {\  (\bibinfo {year}
  {2020})},\ \Eprint {https://arxiv.org/abs/2003.02825} {arXiv:2003.02825
  [quant-ph]} \BibitemShut {NoStop}%
\bibitem [{\citenamefont {{Mark}}\ and\ \citenamefont
  {{Motrunich}}(2020)}]{2020arXiv200413800M}%
  \BibitemOpen
  \bibfield  {author} {\bibinfo {author} {\bibfnamefont {D.~K.}\ \bibnamefont
  {{Mark}}}\ and\ \bibinfo {author} {\bibfnamefont {O.~I.}\ \bibnamefont
  {{Motrunich}}},\ }\bibfield  {title} {\bibinfo {title} {{Eta-pairing states
  as true scars in an extended Hubbard Model}},\ }\href@noop {} {\bibfield
  {journal} {\bibinfo  {journal} {arXiv e-prints}\ ,\ \bibinfo {eid}
  {arXiv:2004.13800}} (\bibinfo {year} {2020})},\ \Eprint
  {https://arxiv.org/abs/2004.13800} {arXiv:2004.13800 [cond-mat.str-el]}
  \BibitemShut {NoStop}%
\bibitem [{\citenamefont {Bull}\ \emph {et~al.}(2019)\citenamefont {Bull},
  \citenamefont {Martin},\ and\ \citenamefont {{Papi\ifmmode \acute{c}\else
  {\'c}\fi{}}}}]{PRLPapicClockModels}%
  \BibitemOpen
  \bibfield  {author} {\bibinfo {author} {\bibfnamefont {K.}~\bibnamefont
  {Bull}}, \bibinfo {author} {\bibfnamefont {I.}~\bibnamefont {Martin}},\ and\
  \bibinfo {author} {\bibfnamefont {Z.}~\bibnamefont {{Papi\ifmmode
  \acute{c}\else {\'c}\fi{}}}},\ }\bibfield  {title} {\bibinfo {title}
  {Systematic construction of scarred many-body dynamics in 1d lattice
  models},\ }\href {https://doi.org/10.1103/PhysRevLett.123.030601} {\bibfield
  {journal} {\bibinfo  {journal} {Phys. Rev. Lett.}\ }\textbf {\bibinfo
  {volume} {123}},\ \bibinfo {pages} {030601} (\bibinfo {year}
  {2019})}\BibitemShut {NoStop}%
\bibitem [{\citenamefont {Khemani}\ \emph {et~al.}(2019)\citenamefont
  {Khemani}, \citenamefont {Laumann},\ and\ \citenamefont
  {Chandran}}]{VedikaScarsVsIntegr}%
  \BibitemOpen
  \bibfield  {author} {\bibinfo {author} {\bibfnamefont {V.}~\bibnamefont
  {Khemani}}, \bibinfo {author} {\bibfnamefont {C.~R.}\ \bibnamefont
  {Laumann}},\ and\ \bibinfo {author} {\bibfnamefont {A.}~\bibnamefont
  {Chandran}},\ }\bibfield  {title} {\bibinfo {title} {{Signatures of
  integrability in the dynamics of Rydberg-blockaded chains}},\ }\href
  {https://doi.org/10.1103/PhysRevB.99.161101} {\bibfield  {journal} {\bibinfo
  {journal} {Phys. Rev. B}\ }\textbf {\bibinfo {volume} {99}},\ \bibinfo
  {pages} {161101} (\bibinfo {year} {2019})}\BibitemShut {NoStop}%
\bibitem [{\citenamefont {Lee}\ \emph {et~al.}(2020)\citenamefont {Lee},
  \citenamefont {Melendrez}, \citenamefont {Pal},\ and\ \citenamefont
  {Changlani}}]{Pal2020ScarsFromFrustration}%
  \BibitemOpen
  \bibfield  {author} {\bibinfo {author} {\bibfnamefont {K.}~\bibnamefont
  {Lee}}, \bibinfo {author} {\bibfnamefont {R.}~\bibnamefont {Melendrez}},
  \bibinfo {author} {\bibfnamefont {A.}~\bibnamefont {Pal}},\ and\ \bibinfo
  {author} {\bibfnamefont {H.~J.}\ \bibnamefont {Changlani}},\ }\bibfield
  {title} {\bibinfo {title} {Exact three-colored quantum scars from geometric
  frustration},\ }\href@noop {} {\bibfield  {journal} {\bibinfo  {journal}
  {Physical Review B}\ }\textbf {\bibinfo {volume} {101}},\ \bibinfo {pages}
  {241111} (\bibinfo {year} {2020})}\BibitemShut {NoStop}%
\bibitem [{\citenamefont {Mark}\ \emph {et~al.}(2020)\citenamefont {Mark},
  \citenamefont {Lin},\ and\ \citenamefont {Motrunich}}]{mark2020unified}%
  \BibitemOpen
  \bibfield  {author} {\bibinfo {author} {\bibfnamefont {D.~K.}\ \bibnamefont
  {Mark}}, \bibinfo {author} {\bibfnamefont {C.-J.}\ \bibnamefont {Lin}},\ and\
  \bibinfo {author} {\bibfnamefont {O.~I.}\ \bibnamefont {Motrunich}},\
  }\bibfield  {title} {\bibinfo {title} {Unified structure for exact towers of
  scar states in the affleck-kennedy-lieb-tasaki and other models},\
  }\href@noop {} {\bibfield  {journal} {\bibinfo  {journal} {Physical Review
  B}\ }\textbf {\bibinfo {volume} {101}},\ \bibinfo {pages} {195131} (\bibinfo
  {year} {2020})}\BibitemShut {NoStop}%
\bibitem [{\citenamefont {Iadecola}\ and\ \citenamefont
  {Schecter}(2020)}]{iadecola2020quantum}%
  \BibitemOpen
  \bibfield  {author} {\bibinfo {author} {\bibfnamefont {T.}~\bibnamefont
  {Iadecola}}\ and\ \bibinfo {author} {\bibfnamefont {M.}~\bibnamefont
  {Schecter}},\ }\bibfield  {title} {\bibinfo {title} {Quantum many-body scar
  states with emergent kinetic constraints and finite-entanglement revivals},\
  }\href@noop {} {\bibfield  {journal} {\bibinfo  {journal} {Physical Review
  B}\ }\textbf {\bibinfo {volume} {101}},\ \bibinfo {pages} {024306} (\bibinfo
  {year} {2020})}\BibitemShut {NoStop}%
\bibitem [{\citenamefont {Moudgalya}\ \emph {et~al.}(2020)\citenamefont
  {Moudgalya}, \citenamefont {Regnault},\ and\ \citenamefont
  {Bernevig}}]{moudgalya2020etapairing}%
  \BibitemOpen
  \bibfield  {author} {\bibinfo {author} {\bibfnamefont {S.}~\bibnamefont
  {Moudgalya}}, \bibinfo {author} {\bibfnamefont {N.}~\bibnamefont
  {Regnault}},\ and\ \bibinfo {author} {\bibfnamefont {B.~A.}\ \bibnamefont
  {Bernevig}},\ }\href@noop {} {\bibinfo {title} {Eta-pairing in hubbard
  models: From spectrum generating algebras to quantum many-body scars}}
  (\bibinfo {year} {2020}),\ \Eprint {https://arxiv.org/abs/2004.13727}
  {arXiv:2004.13727 [cond-mat.str-el]} \BibitemShut {NoStop}%
\bibitem [{\citenamefont {Magnifico}\ \emph {et~al.}(2019)\citenamefont
  {Magnifico}, \citenamefont {Dalmonte}, \citenamefont {Facchi}, \citenamefont
  {Pascazio}, \citenamefont {Pepe},\ and\ \citenamefont
  {Ercolessi}}]{Magnifico:2019kyj}%
  \BibitemOpen
  \bibfield  {author} {\bibinfo {author} {\bibfnamefont {G.}~\bibnamefont
  {Magnifico}}, \bibinfo {author} {\bibfnamefont {M.}~\bibnamefont {Dalmonte}},
  \bibinfo {author} {\bibfnamefont {P.}~\bibnamefont {Facchi}}, \bibinfo
  {author} {\bibfnamefont {S.}~\bibnamefont {Pascazio}}, \bibinfo {author}
  {\bibfnamefont {F.~V.}\ \bibnamefont {Pepe}},\ and\ \bibinfo {author}
  {\bibfnamefont {E.}~\bibnamefont {Ercolessi}},\ }\bibfield  {title} {\bibinfo
  {title} {{Real Time Dynamics and Confinement in the $\mathbb{Z}_{n}$
  Schwinger-Weyl lattice model for 1+1 QED}}\ }\href
  {https://doi.org/10.22331/q-2020-06-15-281} {10.22331/q-2020-06-15-281}
  (\bibinfo {year} {2019}),\ \Eprint {https://arxiv.org/abs/1909.04821}
  {arXiv:1909.04821 [quant-ph]} \BibitemShut {NoStop}%
\bibitem [{\citenamefont {Yang}(1989)}]{etaPairingYang89}%
  \BibitemOpen
  \bibfield  {author} {\bibinfo {author} {\bibfnamefont {C.~N.}\ \bibnamefont
  {Yang}},\ }\bibfield  {title} {\bibinfo {title} {\ensuremath{\eta} pairing
  and off-diagonal long-range order in a hubbard model},\ }\href
  {https://doi.org/10.1103/PhysRevLett.63.2144} {\bibfield  {journal} {\bibinfo
   {journal} {Phys. Rev. Lett.}\ }\textbf {\bibinfo {volume} {63}},\ \bibinfo
  {pages} {2144} (\bibinfo {year} {1989})}\BibitemShut {NoStop}%
\bibitem [{\citenamefont {Yang}\ and\ \citenamefont
  {Zhang}(1990)}]{yang1990so}%
  \BibitemOpen
  \bibfield  {author} {\bibinfo {author} {\bibfnamefont {C.~N.}\ \bibnamefont
  {Yang}}\ and\ \bibinfo {author} {\bibfnamefont {S.}~\bibnamefont {Zhang}},\
  }\bibfield  {title} {\bibinfo {title} {So(4) symmetry in a hubbard model},\
  }\href@noop {} {\bibfield  {journal} {\bibinfo  {journal} {Modern Physics
  Letters B}\ }\textbf {\bibinfo {volume} {4}},\ \bibinfo {pages} {759}
  (\bibinfo {year} {1990})}\BibitemShut {NoStop}%
\bibitem [{\citenamefont {Zhang}(1991)}]{ZhangHubbardSO41991}%
  \BibitemOpen
  \bibfield  {author} {\bibinfo {author} {\bibfnamefont {S.}~\bibnamefont
  {Zhang}},\ }\bibfield  {title} {\bibinfo {title} {So(4) symmetry of the
  hubbard model and its experimental consequences},\ }\href
  {https://doi.org/10.1142/S0217979291000110} {\bibfield  {journal} {\bibinfo
  {journal} {International Journal of Modern Physics B}\ }\textbf {\bibinfo
  {volume} {05}},\ \bibinfo {pages} {153} (\bibinfo {year} {1991})},\ \Eprint
  {https://arxiv.org/abs/https://doi.org/10.1142/S0217979291000110}
  {https://doi.org/10.1142/S0217979291000110} \BibitemShut {NoStop}%
\bibitem [{\citenamefont {Bernien}\ \emph {et~al.}(2017)\citenamefont
  {Bernien}, \citenamefont {Schwartz}, \citenamefont {Keesling}, \citenamefont
  {Levine}, \citenamefont {Omran}, \citenamefont {Pichler}, \citenamefont
  {Choi}, \citenamefont {Zibrov}, \citenamefont {Endres}, \citenamefont
  {Greiner}, \citenamefont {Vuleti{\'c}},\ and\ \citenamefont
  {Lukin}}]{RydbergExperimentRevivals}%
  \BibitemOpen
  \bibfield  {author} {\bibinfo {author} {\bibfnamefont {H.}~\bibnamefont
  {Bernien}}, \bibinfo {author} {\bibfnamefont {S.}~\bibnamefont {Schwartz}},
  \bibinfo {author} {\bibfnamefont {A.}~\bibnamefont {Keesling}}, \bibinfo
  {author} {\bibfnamefont {H.}~\bibnamefont {Levine}}, \bibinfo {author}
  {\bibfnamefont {A.}~\bibnamefont {Omran}}, \bibinfo {author} {\bibfnamefont
  {H.}~\bibnamefont {Pichler}}, \bibinfo {author} {\bibfnamefont
  {S.}~\bibnamefont {Choi}}, \bibinfo {author} {\bibfnamefont {A.~S.}\
  \bibnamefont {Zibrov}}, \bibinfo {author} {\bibfnamefont {M.}~\bibnamefont
  {Endres}}, \bibinfo {author} {\bibfnamefont {M.}~\bibnamefont {Greiner}},
  \bibinfo {author} {\bibfnamefont {V.}~\bibnamefont {Vuleti{\'c}}},\ and\
  \bibinfo {author} {\bibfnamefont {M.~D.}\ \bibnamefont {Lukin}},\ }\bibfield
  {title} {\bibinfo {title} {Probing many-body dynamics on a 51-atom quantum
  simulator},\ }\href {https://doi.org/10.1038/nature24622} {\bibfield
  {journal} {\bibinfo  {journal} {Nature}\ }\textbf {\bibinfo {volume} {551}},\
  \bibinfo {pages} {579 EP} (\bibinfo {year} {2017})}\BibitemShut {NoStop}%
\bibitem [{\citenamefont {Deutsch}(1991)}]{deutsch1991quantum}%
  \BibitemOpen
  \bibfield  {author} {\bibinfo {author} {\bibfnamefont {J.~M.}\ \bibnamefont
  {Deutsch}},\ }\bibfield  {title} {\bibinfo {title} {Quantum statistical
  mechanics in a closed system},\ }\href@noop {} {\bibfield  {journal}
  {\bibinfo  {journal} {Physical Review A}\ }\textbf {\bibinfo {volume} {43}},\
  \bibinfo {pages} {2046} (\bibinfo {year} {1991})}\BibitemShut {NoStop}%
\bibitem [{\citenamefont {Srednicki}(1994)}]{srednicki1994chaos}%
  \BibitemOpen
  \bibfield  {author} {\bibinfo {author} {\bibfnamefont {M.}~\bibnamefont
  {Srednicki}},\ }\bibfield  {title} {\bibinfo {title} {Chaos and quantum
  thermalization},\ }\href@noop {} {\bibfield  {journal} {\bibinfo  {journal}
  {Physical Review E}\ }\textbf {\bibinfo {volume} {50}},\ \bibinfo {pages}
  {888} (\bibinfo {year} {1994})}\BibitemShut {NoStop}%
\bibitem [{\citenamefont {Rigol}\ \emph {et~al.}(2008)\citenamefont {Rigol},
  \citenamefont {Dunjko},\ and\ \citenamefont
  {Olshanii}}]{rigol2008thermalization}%
  \BibitemOpen
  \bibfield  {author} {\bibinfo {author} {\bibfnamefont {M.}~\bibnamefont
  {Rigol}}, \bibinfo {author} {\bibfnamefont {V.}~\bibnamefont {Dunjko}},\ and\
  \bibinfo {author} {\bibfnamefont {M.}~\bibnamefont {Olshanii}},\ }\bibfield
  {title} {\bibinfo {title} {Thermalization and its mechanism for generic
  isolated quantum systems},\ }\href@noop {} {\bibfield  {journal} {\bibinfo
  {journal} {Nature}\ }\textbf {\bibinfo {volume} {452}},\ \bibinfo {pages}
  {854} (\bibinfo {year} {2008})}\BibitemShut {NoStop}%
\bibitem [{Note1()}]{Note1}%
  \BibitemOpen
  \bibinfo {note} {This happens, for example, when the energies of all states
  in $\protect \mathbb {S}$ are integers in some units}\BibitemShut {NoStop}%
\bibitem [{\citenamefont {Wu}\ \emph {et~al.}(2019)\citenamefont {Wu},
  \citenamefont {Jian},\ and\ \citenamefont {Xu}}]{PhysRevB.100.075101}%
  \BibitemOpen
  \bibfield  {author} {\bibinfo {author} {\bibfnamefont {X.-C.}\ \bibnamefont
  {Wu}}, \bibinfo {author} {\bibfnamefont {C.-M.}\ \bibnamefont {Jian}},\ and\
  \bibinfo {author} {\bibfnamefont {C.}~\bibnamefont {Xu}},\ }\bibfield
  {title} {\bibinfo {title} {{Lattice models for non-Fermi liquids with tunable
  transport scalings}},\ }\href {https://doi.org/10.1103/PhysRevB.100.075101}
  {\bibfield  {journal} {\bibinfo  {journal} {Phys. Rev. B}\ }\textbf {\bibinfo
  {volume} {100}},\ \bibinfo {pages} {075101} (\bibinfo {year}
  {2019})}\BibitemShut {NoStop}%
\bibitem [{\citenamefont {Klebanov}\ \emph
  {et~al.}(2018{\natexlab{a}})\citenamefont {Klebanov}, \citenamefont
  {Milekhin}, \citenamefont {Popov},\ and\ \citenamefont
  {Tarnopolsky}}]{Klebanov:2018nfp}%
  \BibitemOpen
  \bibfield  {author} {\bibinfo {author} {\bibfnamefont {I.~R.}\ \bibnamefont
  {Klebanov}}, \bibinfo {author} {\bibfnamefont {A.}~\bibnamefont {Milekhin}},
  \bibinfo {author} {\bibfnamefont {F.}~\bibnamefont {Popov}},\ and\ \bibinfo
  {author} {\bibfnamefont {G.}~\bibnamefont {Tarnopolsky}},\ }\bibfield
  {title} {\bibinfo {title} {{Spectra of eigenstates in fermionic tensor
  quantum mechanics}},\ }\href {https://doi.org/10.1103/PhysRevD.97.106023}
  {\bibfield  {journal} {\bibinfo  {journal} {Phys. Rev.}\ }\textbf {\bibinfo
  {volume} {D97}},\ \bibinfo {pages} {106023} (\bibinfo {year}
  {2018}{\natexlab{a}})},\ \Eprint {https://arxiv.org/abs/1802.10263}
  {arXiv:1802.10263 [hep-th]} \BibitemShut {NoStop}%
\bibitem [{\citenamefont {Gaitan}\ \emph {et~al.}(2020)\citenamefont {Gaitan},
  \citenamefont {Klebanov}, \citenamefont {Pakrouski}, \citenamefont
  {Pallegar},\ and\ \citenamefont {Popov}}]{Gaitan:2020zbm}%
  \BibitemOpen
  \bibfield  {author} {\bibinfo {author} {\bibfnamefont {G.}~\bibnamefont
  {Gaitan}}, \bibinfo {author} {\bibfnamefont {I.~R.}\ \bibnamefont
  {Klebanov}}, \bibinfo {author} {\bibfnamefont {K.}~\bibnamefont {Pakrouski}},
  \bibinfo {author} {\bibfnamefont {P.~N.}\ \bibnamefont {Pallegar}},\ and\
  \bibinfo {author} {\bibfnamefont {F.~K.}\ \bibnamefont {Popov}},\ }\bibfield
  {title} {\bibinfo {title} {{Hagedorn Temperature in Large $N$ Majorana
  Quantum Mechanics}},\ }\href {https://doi.org/10.1103/PhysRevD.101.126002}
  {\bibfield  {journal} {\bibinfo  {journal} {Phys. Rev. D}\ }\textbf {\bibinfo
  {volume} {101}},\ \bibinfo {pages} {126002} (\bibinfo {year} {2020})},\
  \Eprint {https://arxiv.org/abs/2002.02066} {arXiv:2002.02066 [hep-th]}
  \BibitemShut {NoStop}%
\bibitem [{Note2()}]{Note2}%
  \BibitemOpen
  \bibinfo {note} {A more general version of $H_0$ \cite {Klebanov:2018fzb} is
  considered in Sec. I of the SM where the choice of a smaller group $G$ leads
  to the dimension of the scar subsector growing exponentially in
  $M=N_1N_2$.}\BibitemShut {Stop}%
\bibitem [{\citenamefont {Klebanov}\ and\ \citenamefont
  {Tarnopolsky}(2017)}]{Klebanov:2016xxf}%
  \BibitemOpen
  \bibfield  {author} {\bibinfo {author} {\bibfnamefont {I.~R.}\ \bibnamefont
  {Klebanov}}\ and\ \bibinfo {author} {\bibfnamefont {G.}~\bibnamefont
  {Tarnopolsky}},\ }\bibfield  {title} {\bibinfo {title} {{Uncolored random
  tensors, melon diagrams, and the Sachdev-Ye-Kitaev models}},\ }\href
  {https://doi.org/10.1103/PhysRevD.95.046004} {\bibfield  {journal} {\bibinfo
  {journal} {Phys. Rev.}\ }\textbf {\bibinfo {volume} {D95}},\ \bibinfo {pages}
  {046004} (\bibinfo {year} {2017})},\ \Eprint
  {https://arxiv.org/abs/1611.08915} {arXiv:1611.08915 [hep-th]} \BibitemShut
  {NoStop}%
\bibitem [{\citenamefont {Pakrouski}\ \emph {et~al.}(2019)\citenamefont
  {Pakrouski}, \citenamefont {Klebanov}, \citenamefont {Popov},\ and\
  \citenamefont {Tarnopolsky}}]{Pakrouski:2018jcc}%
  \BibitemOpen
  \bibfield  {author} {\bibinfo {author} {\bibfnamefont {K.}~\bibnamefont
  {Pakrouski}}, \bibinfo {author} {\bibfnamefont {I.~R.}\ \bibnamefont
  {Klebanov}}, \bibinfo {author} {\bibfnamefont {F.}~\bibnamefont {Popov}},\
  and\ \bibinfo {author} {\bibfnamefont {G.}~\bibnamefont {Tarnopolsky}},\
  }\bibfield  {title} {\bibinfo {title} {{Spectrum of Majorana Quantum
  Mechanics with $O(4)^3$ Symmetry}},\ }\href
  {https://doi.org/10.1103/PhysRevLett.122.011601} {\bibfield  {journal}
  {\bibinfo  {journal} {Phys. Rev. Lett.}\ }\textbf {\bibinfo {volume} {122}},\
  \bibinfo {pages} {011601} (\bibinfo {year} {2019})},\ \Eprint
  {https://arxiv.org/abs/1808.07455} {arXiv:1808.07455 [hep-th]} \BibitemShut
  {NoStop}%
\bibitem [{Note3()}]{Note3}%
  \BibitemOpen
  \bibinfo {note} {The Hamiltonian $H=H_0+OT$ has the same properties with
  respect to the presence of the many-body scar states.}\BibitemShut {Stop}%
\bibitem [{\citenamefont {Cotler}\ \emph {et~al.}(2017)\citenamefont {Cotler},
  \citenamefont {Gur-Ari}, \citenamefont {Hanada}, \citenamefont {Polchinski},
  \citenamefont {Saad}, \citenamefont {Shenker}, \citenamefont {Stanford},
  \citenamefont {Streicher},\ and\ \citenamefont {Tezuka}}]{Cotler:2016fpe}%
  \BibitemOpen
  \bibfield  {author} {\bibinfo {author} {\bibfnamefont {J.~S.}\ \bibnamefont
  {Cotler}}, \bibinfo {author} {\bibfnamefont {G.}~\bibnamefont {Gur-Ari}},
  \bibinfo {author} {\bibfnamefont {M.}~\bibnamefont {Hanada}}, \bibinfo
  {author} {\bibfnamefont {J.}~\bibnamefont {Polchinski}}, \bibinfo {author}
  {\bibfnamefont {P.}~\bibnamefont {Saad}}, \bibinfo {author} {\bibfnamefont
  {S.~H.}\ \bibnamefont {Shenker}}, \bibinfo {author} {\bibfnamefont
  {D.}~\bibnamefont {Stanford}}, \bibinfo {author} {\bibfnamefont
  {A.}~\bibnamefont {Streicher}},\ and\ \bibinfo {author} {\bibfnamefont
  {M.}~\bibnamefont {Tezuka}},\ }\bibfield  {title} {\bibinfo {title} {{Black
  Holes and Random Matrices}},\ }\href
  {https://doi.org/10.1007/JHEP05(2017)118} {\bibfield  {journal} {\bibinfo
  {journal} {JHEP}\ }\textbf {\bibinfo {volume} {05}},\ \bibinfo {pages}
  {118}},\ \bibinfo {note} {[Erratum: JHEP 09, 002 (2018)]},\ \Eprint
  {https://arxiv.org/abs/1611.04650} {arXiv:1611.04650 [hep-th]} \BibitemShut
  {NoStop}%
\bibitem [{\citenamefont {Bu{\v c}a}\ \emph {et~al.}(2019)\citenamefont {Bu{\v
  c}a}, \citenamefont {Tindall},\ and\ \citenamefont
  {Jaksch}}]{BucaNature2019}%
  \BibitemOpen
  \bibfield  {author} {\bibinfo {author} {\bibfnamefont {B.}~\bibnamefont
  {Bu{\v c}a}}, \bibinfo {author} {\bibfnamefont {J.}~\bibnamefont {Tindall}},\
  and\ \bibinfo {author} {\bibfnamefont {D.}~\bibnamefont {Jaksch}},\
  }\bibfield  {title} {\bibinfo {title} {Non-stationary coherent quantum
  many-body dynamics through dissipation},\ }\href
  {https://doi.org/10.1038/s41467-019-09757-y} {\bibfield  {journal} {\bibinfo
  {journal} {Nature Communications}\ }\textbf {\bibinfo {volume} {10}},\
  \bibinfo {pages} {1730} (\bibinfo {year} {2019})}\BibitemShut {NoStop}%
\bibitem [{\citenamefont {Yang}(1962)}]{yang1962concept}%
  \BibitemOpen
  \bibfield  {author} {\bibinfo {author} {\bibfnamefont {C.~N.}\ \bibnamefont
  {Yang}},\ }\bibfield  {title} {\bibinfo {title} {Concept of off-diagonal
  long-range order and the quantum phases of liquid he and of
  superconductors},\ }\href@noop {} {\bibfield  {journal} {\bibinfo  {journal}
  {Reviews of Modern Physics}\ }\textbf {\bibinfo {volume} {34}},\ \bibinfo
  {pages} {694} (\bibinfo {year} {1962})}\BibitemShut {NoStop}%
\bibitem [{\citenamefont {Milekhin}(2020)}]{Milekhin:2020zpg}%
  \BibitemOpen
  \bibfield  {author} {\bibinfo {author} {\bibfnamefont {A.}~\bibnamefont
  {Milekhin}},\ }\bibfield  {title} {\bibinfo {title} {{Quantum error
  correction and large $N$}},\ }\href@noop {} {\  (\bibinfo {year} {2020})},\
  \Eprint {https://arxiv.org/abs/2008.12869} {arXiv:2008.12869 [hep-th]}
  \BibitemShut {NoStop}%
\bibitem [{\citenamefont {Lidar}\ \emph {et~al.}(1998)\citenamefont {Lidar},
  \citenamefont {Chuang},\ and\ \citenamefont
  {Whaley}}]{DecohFreeQComputation1998Chuang}%
  \BibitemOpen
  \bibfield  {author} {\bibinfo {author} {\bibfnamefont {D.~A.}\ \bibnamefont
  {Lidar}}, \bibinfo {author} {\bibfnamefont {I.~L.}\ \bibnamefont {Chuang}},\
  and\ \bibinfo {author} {\bibfnamefont {K.~B.}\ \bibnamefont {Whaley}},\
  }\bibfield  {title} {\bibinfo {title} {Decoherence-free subspaces for quantum
  computation},\ }\href {https://doi.org/10.1103/PhysRevLett.81.2594}
  {\bibfield  {journal} {\bibinfo  {journal} {Phys. Rev. Lett.}\ }\textbf
  {\bibinfo {volume} {81}},\ \bibinfo {pages} {2594} (\bibinfo {year}
  {1998})}\BibitemShut {NoStop}%
\bibitem [{\citenamefont {Maldacena}(1999)}]{Maldacena:1997re}%
  \BibitemOpen
  \bibfield  {author} {\bibinfo {author} {\bibfnamefont {J.~M.}\ \bibnamefont
  {Maldacena}},\ }\bibfield  {title} {\bibinfo {title} {{The Large N limit of
  superconformal field theories and supergravity}},\ }\href
  {https://doi.org/10.1023/A:1026654312961; 10.4310/ATMP.1998.v2.n2.a1}
  {\bibfield  {journal} {\bibinfo  {journal} {Int. J. Theor. Phys.}\ }\textbf
  {\bibinfo {volume} {38}},\ \bibinfo {pages} {1113} (\bibinfo {year}
  {1999})},\ \bibinfo {note} {[Adv. Theor. Math. Phys.2,231(1998)]},\ \Eprint
  {https://arxiv.org/abs/hep-th/9711200} {arXiv:hep-th/9711200 [hep-th]}
  \BibitemShut {NoStop}%
\bibitem [{\citenamefont {Gubser}\ \emph {et~al.}(1998)\citenamefont {Gubser},
  \citenamefont {Klebanov},\ and\ \citenamefont {Polyakov}}]{Gubser:1998bc}%
  \BibitemOpen
  \bibfield  {author} {\bibinfo {author} {\bibfnamefont {S.~S.}\ \bibnamefont
  {Gubser}}, \bibinfo {author} {\bibfnamefont {I.~R.}\ \bibnamefont
  {Klebanov}},\ and\ \bibinfo {author} {\bibfnamefont {A.~M.}\ \bibnamefont
  {Polyakov}},\ }\bibfield  {title} {\bibinfo {title} {{Gauge theory
  correlators from noncritical string theory}},\ }\href
  {https://doi.org/10.1016/S0370-2693(98)00377-3} {\bibfield  {journal}
  {\bibinfo  {journal} {Phys. Lett.}\ }\textbf {\bibinfo {volume} {B428}},\
  \bibinfo {pages} {105} (\bibinfo {year} {1998})},\ \Eprint
  {https://arxiv.org/abs/hep-th/9802109} {arXiv:hep-th/9802109 [hep-th]}
  \BibitemShut {NoStop}%
\bibitem [{\citenamefont {Witten}(1998)}]{Witten:1998qj}%
  \BibitemOpen
  \bibfield  {author} {\bibinfo {author} {\bibfnamefont {E.}~\bibnamefont
  {Witten}},\ }\bibfield  {title} {\bibinfo {title} {{Anti-de Sitter space and
  holography}},\ }\href {https://doi.org/10.4310/ATMP.1998.v2.n2.a2} {\bibfield
   {journal} {\bibinfo  {journal} {Adv. Theor. Math. Phys.}\ }\textbf {\bibinfo
  {volume} {2}},\ \bibinfo {pages} {253} (\bibinfo {year} {1998})},\ \Eprint
  {https://arxiv.org/abs/hep-th/9802150} {arXiv:hep-th/9802150 [hep-th]}
  \BibitemShut {NoStop}%
\bibitem [{\citenamefont {Gross}\ and\ \citenamefont
  {Klebanov}(1990)}]{Gross:1990ub}%
  \BibitemOpen
  \bibfield  {author} {\bibinfo {author} {\bibfnamefont {D.~J.}\ \bibnamefont
  {Gross}}\ and\ \bibinfo {author} {\bibfnamefont {I.~R.}\ \bibnamefont
  {Klebanov}},\ }\bibfield  {title} {\bibinfo {title} {{One-dimensional string
  theory on a circle}},\ }\href {https://doi.org/10.1016/0550-3213(90)90667-3}
  {\bibfield  {journal} {\bibinfo  {journal} {Nucl. Phys. B}\ }\textbf
  {\bibinfo {volume} {344}},\ \bibinfo {pages} {475} (\bibinfo {year}
  {1990})}\BibitemShut {NoStop}%
\bibitem [{\citenamefont {Gross}\ and\ \citenamefont
  {Klebanov}(1991)}]{Gross:1990md}%
  \BibitemOpen
  \bibfield  {author} {\bibinfo {author} {\bibfnamefont {D.~J.}\ \bibnamefont
  {Gross}}\ and\ \bibinfo {author} {\bibfnamefont {I.~R.}\ \bibnamefont
  {Klebanov}},\ }\bibfield  {title} {\bibinfo {title} {{Vortices and the
  nonsinglet sector of the $c = 1$ matrix model}},\ }\href
  {https://doi.org/10.1016/0550-3213(91)90363-3} {\bibfield  {journal}
  {\bibinfo  {journal} {Nucl. Phys. B}\ }\textbf {\bibinfo {volume} {354}},\
  \bibinfo {pages} {459} (\bibinfo {year} {1991})}\BibitemShut {NoStop}%
\bibitem [{\citenamefont {Witten}(2019)}]{Witten:2016iux}%
  \BibitemOpen
  \bibfield  {author} {\bibinfo {author} {\bibfnamefont {E.}~\bibnamefont
  {Witten}},\ }\bibfield  {title} {\bibinfo {title} {{An SYK-Like Model Without
  Disorder}},\ }\href {https://doi.org/10.1088/1751-8121/ab3752} {\bibfield
  {journal} {\bibinfo  {journal} {J. Phys.}\ }\textbf {\bibinfo {volume}
  {A52}},\ \bibinfo {pages} {474002} (\bibinfo {year} {2019})},\ \Eprint
  {https://arxiv.org/abs/1610.09758} {arXiv:1610.09758 [hep-th]} \BibitemShut
  {NoStop}%
\bibitem [{\citenamefont {Maldacena}\ and\ \citenamefont
  {Milekhin}(2018)}]{Maldacena:2018vsr}%
  \BibitemOpen
  \bibfield  {author} {\bibinfo {author} {\bibfnamefont {J.}~\bibnamefont
  {Maldacena}}\ and\ \bibinfo {author} {\bibfnamefont {A.}~\bibnamefont
  {Milekhin}},\ }\bibfield  {title} {\bibinfo {title} {{To gauge or not to
  gauge?}},\ }\href {https://doi.org/10.1007/JHEP04(2018)084} {\bibfield
  {journal} {\bibinfo  {journal} {JHEP}\ }\textbf {\bibinfo {volume} {04}},\
  \bibinfo {pages} {084}},\ \Eprint {https://arxiv.org/abs/1802.00428}
  {arXiv:1802.00428 [hep-th]} \BibitemShut {NoStop}%
\bibitem [{\citenamefont {Klebanov}\ \emph
  {et~al.}(2018{\natexlab{b}})\citenamefont {Klebanov}, \citenamefont {Popov},\
  and\ \citenamefont {Tarnopolsky}}]{Klebanov:2018fzb}%
  \BibitemOpen
  \bibfield  {author} {\bibinfo {author} {\bibfnamefont {I.~R.}\ \bibnamefont
  {Klebanov}}, \bibinfo {author} {\bibfnamefont {F.}~\bibnamefont {Popov}},\
  and\ \bibinfo {author} {\bibfnamefont {G.}~\bibnamefont {Tarnopolsky}},\
  }\bibfield  {title} {\bibinfo {title} {{TASI Lectures on Large $N$ Tensor
  Models}},\ }\bibfield  {booktitle} {\emph {\bibinfo {booktitle}
  {{Proceedings, Theoretical Advanced Study Institute in Elementary Particle
  Physics: Physics at the Fundamental Frontier (TASI 2017): Boulder, CO, USA,
  June 5-30, 2017}}},\ }\href {https://doi.org/10.22323/1.305.0004} {\bibfield
  {journal} {\bibinfo  {journal} {PoS}\ }\textbf {\bibinfo {volume}
  {TASI2017}},\ \bibinfo {pages} {004} (\bibinfo {year}
  {2018}{\natexlab{b}})},\ \Eprint {https://arxiv.org/abs/1808.09434}
  {arXiv:1808.09434 [hep-th]} \BibitemShut {NoStop}%
\end{thebibliography}%
\end{document}